**Continuous-distribution puddle model for conduction in trilayer graphene**

Richard S. Thompson[a], Yi- Chen Chang, and Jia G. Lu,

Department of Physics & Astronomy, University of Southern California, Los Angeles,

California 90089-0484, USA

**Abstract.** The temperature dependence of the resistance in trilayer graphene is observed under different applied gate voltages. At small gate voltages the resistance decreases with increasing temperature due to the increase in carrier concentration resulting from thermal excitation of electron-hole pairs, characteristic of a semimetal. At large gate voltages excitation of electron-hole pairs is suppressed, and the resistance increases with increasing temperature because of the enhanced electron-phonon scattering, characteristic of a metal. We find that the simple model with overlapping conduction and valence bands, each with quadratic dispersion relations, is unsatisfactory. Instead, we conclude that impurities in the substrate that create local puddles of higher electron or hole densities are responsible for the residual conductivity at low temperatures. The best fit is obtained using a continuous distribution of puddles. From the fit the average of the electron and hole effective masses can be determined.

# 1 Introduction

Since the pioneering work of Novoselov *et al*. [1] extensive work has been done on monolayer and bilayer graphene. A number of review articles have appeared, including Castro Neto *et al*. [2], Peres [3], and Das Sarma *et al*. [4]. However, not much work has been done on few-layer graphene. An early work on trilayer graphene was done by Craciun *et al*. [5]. They interpreted the residual conductivity at low temperatures as a result of band overlap. However, from our experimental study of the conductance of trilayer graphene we find that this model does not give a good fit at low temperatures and that the inferred band overlap is twice as large as their value, indicating that it is not a universal quantity but varies from sample to sample. More recently, scanning single-electron tunneling (SET) [6] and scanning-tunneling microscope (STM) [7,8] measurements have shown that the density of electrons is not constant for graphene films mounted on $SiO_2$ substrates. There are local puddles of higher electron or hole densities caused by impurities in the substrate. Thus we have analyzed our data using the puddle model and determined a good fit to our data using a continuous distribution of puddles.

# 2 Experiment

Few-layer graphene samples were extrapolated from highly oriented pyrolygic graphite (HOPG) by the peeling off process as described in reference [1]. After several peeling processes, micron-sized few-layer graphene samples with thicknesses ranging from 1 to 4 nm were fabricated and then transferred onto a $SiO_2$ (300 nm)/Si substrate with alignment marks that were patterned via photolithography. Sheets of few-layer graphene were located by an optical microscope. From atomic force microscopy (AFM), the sample thickness was estimated to be 1.5 nm, which corresponds to the thickness of between four and five layers of graphene with the interlayer spacing of 0.335 nm. However, this thickness measurement was not very accurate, since it could include a dead layer or a layer of water. Raman scattering shows that the sample has more than two layers but cannot conclude the exact number. The actual thickness of our sample was determined to be three layers (shown later).

Standard electron-beam lithography was adopted to fabricate electrodes. Pd (10 nm)/Au (60nm) contacts were deposited by an electron-beam sputter after development. The resistance was measured as a function of temperature and of the voltage applied to a gate electrode. The resistivity measurements were carried out in a cryostat over a temperature range from 4.5 to 300 K.

For our measurements of the electrostatic-field effect in few-layer graphene we used a sample with width $w = 1.15\,\mu\mathrm{m}$ and length $L = 4.3\,\mu\mathrm{m}$ as shown in the inset of Figure 1. Because of the irregular shape of our sample the value of the width is approximate, but this does not affect the temperature and gate voltage dependence of our measurements. Our sample was slightly p-doped due to absorption of humidity [1,9].

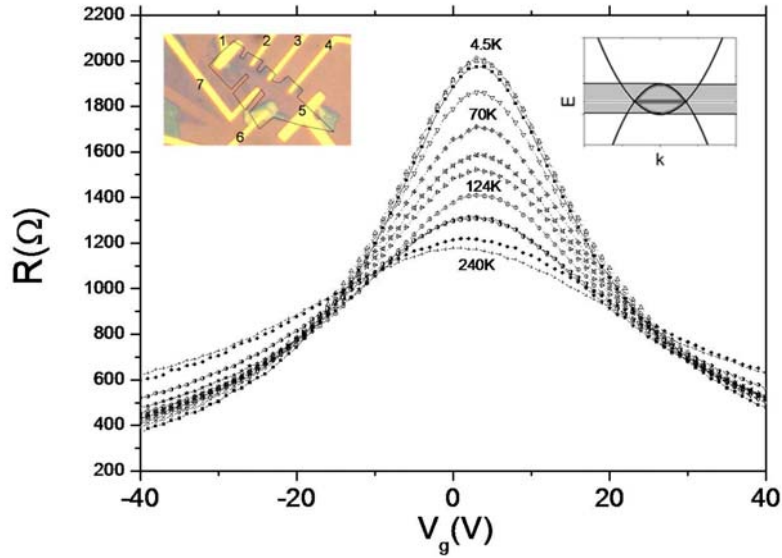

FIG. 1. (Color online) Measured resistance of trilayer graphene as a function of the gate voltage $V_g$ at different temperatures (4.5, 10, 30, 50, 70, 90, 100, 124, 155, 170, 220, and 240 K, respectively). The upper left inset shows an optical image of the device. The current is injected between probes 1 and 4, and the voltage is measured between probes 2 and 3. The distance between probes 2 and 3 is $L = 4.3\,\mu\mathrm{m}$. The upper right insert shows the band structure for the simple two-band model with a band overlap (shaded region) of width $E_0$.

The results of our resistance measurements are plotted in Figure 1 as a function of the gate voltage $V_g$ at different temperatures. Figure 2 shows the same data as a function of temperature under different gate voltages.

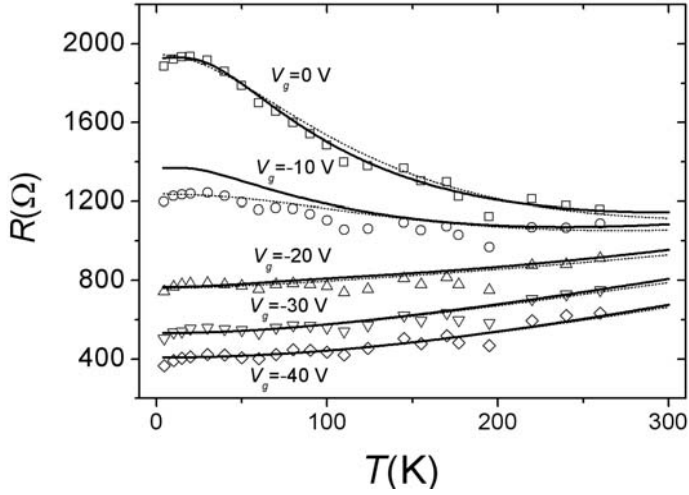

FIG. 2. Temperature dependence of the resistance at several gate voltages ranging from -40V to 0 V. The solid lines are fits to the overlapping-band theory, while the dotted lines are fits to the continuous-distribution puddle model.

We observe that the resistance decreases with increasing temperature for small gate voltage, while it increases with temperature for large gate voltage. For small gate voltages the number of carriers is small at low temperatures, and thermal excitation of electron-hole pairs gives rise to a rapid increase in the carrier concentration and therefore in the conductivity. When the gate potential is large the density of carriers is almost independent of temperature, and the resistance increases with temperature due to the decrease in the mobility caused by enhanced electron-phonon scattering. This behavior can be called a semimetal-to-metal transition due to the different fillings of the band with electrons when the chemical potential is shifted by the gate voltage. It is also called an insulator-to-metal transition in the literature [10] when there is no measurable band overlap or gap but the residual conductivity at low temperature is due to spatial inhomogeneity (puddles) in the sample. We don't know the cause of the small downturn of the resistance at the lowest temperature, but it could be due to a heating effect.

The mobilities $\mu_e$ of the electrons and $\mu_h$ of the holes are estimated from the slope of the conductivity versus gate voltage for large positive or negative values. In these limits the conductivity is dominated by only one type of carrier because the chemical potential has been shifted far out of the band overlap region shown in the upper right inset of Figure 1. The two-dimensional conductivity (conductance per square) is given by $\sigma = e(n_e \mu_e + n_h \mu_h)$, where $n_e$ is the density of electrons and $n_h$ is the density of holes per unit area.   At large positive gate voltages the holes have all been filled, and the number of electrons in the system is given by $CV_g/e$, where $C$ is the capacitance between the gate electrode and the sample.   The capacitance per area is $C/A = \varepsilon_0 \varepsilon_r / d$, with $\varepsilon_r = 3.9$ for the SiO$_2$ dielectric and the separation between the sample and the gate electrode $d = 300\,\text{nm}$.   At large positive values of $V_g$ the limiting equation for the conductivity is $\sigma = \mu_e C V_g / A$, and we can estimate $\mu_e$ from the slope $d\sigma/dV_g$.   This estimate for the mobility is then adjusted to give the best fit to the data over the whole range of gate voltages.   A similar method is used to find the hole mobility $\mu_e$ from the conductivity at large negative $V_g$.   The results for the mobilities are plotted in Figure 3.

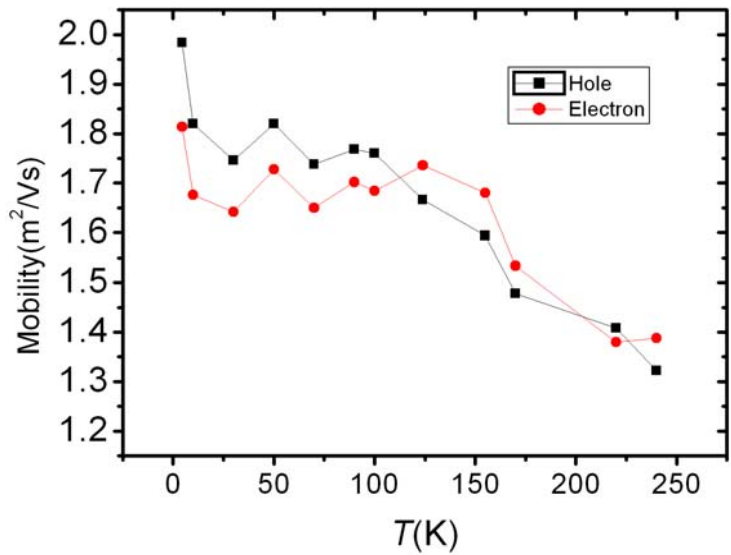

FIG. 3. (Color online) The mobilities $\mu_e$ for electrons and $\mu_h$ for holes as functions of temperature.

Both mobilities are fit well by an expression of the empirical functional form

$\mu(T) = \mu_0 / (1 + aT^2)$. The exponent of the temperature dependence is chosen to give the best fit for integer values. The two mobilities differ from each other by less than 10%. This mobility result could imply that the effective masses of the electrons and holes are approximately equal. However, it is known experimentally that the holes are about 30% to 40% heavier than the electrons in bilayer graphene [11,12]. We don't know why the mobilities are more nearly equal than the effective masses. Nevertheless, we will use a model in which the two effective masses are taken to be the same, as was done in reference [5], since this allows us to analyze the dependence of our experimental results on gate voltage analytically.

**3 Theory**

The band structure of few-layer graphene has been calculated by Partoens and Peeters [13] and by Grüneis, *et al.* [14] in a third-nearest-neighbor tight binding (TB) formalism. For half filling, one free electron per carbon atom, the important states near the Fermi energy $E_F$ are located near one of the corners of the hexagonal reciprocal lattice, the K points. There are two inequivalent K points in the first Brillouin zone. Energy is measured relative to the Fermi energy ($E_F = 0$). For single-layered graphene there are two Dirac bands with a linear energy-momentum dispersion relation, $E = \pm vp$, where *p* is the magnitude of the deviation of the momentum from its value at the K point and *v* denotes the Fermi velocity. The two bands touch at the K point where $E = 0$. For two-layer graphene the picture changes completely. There are four bands, two conduction bands and two valence bands that all have quadratic dispersion relations. Two of the bands touch at the K point, like for the single-layer case, while the other two bands are split away from $E = 0$ by a gap. For three-layer graphene there are six bands, a combination of the bands for the one- and two-layer case. For four-layer graphene there are eight bands, all with quadratic dispersion. Half of the bands approach $E = 0$ at the K point, while the other half are split off by gaps. The alternation of Dirac bands with quadratic bands for odd-numbered multilayers and of only quadratic bands for even-numbered multilayers persists for higher numbers of layers. This band structure has been observed experimentally by Ohta *et.al.* [15] using angle-resolved photoemission spectroscopy (ARPES).

The main features of this energy band structure can be obtained from a simple

TB model that considers only nearest neighbor interactions within the layers and between layers [2,16]. The lattice structure of single-layer graphene is a honeycomb in the $x$-$y$ plane with two inequivalent nearest neighbor sites ($A$ and $B$) in the unit cell. The TB Hamiltonian for the two sublattices has the form

$$H = \begin{pmatrix} 0 & v\pi \\ v\pi^* & 0 \end{pmatrix} \quad (1)$$

where $\pi = p_x + ip_y$ and $v = 3\gamma_0 a/2\hbar$. $\gamma_0 = 3\,\mathrm{eV}$ is the intralayer hopping energy, and $a = 1.42\,\mathrm{\AA}$ is the intralayer nearest neighbor distance, giving $v = 1.0\times 10^6\,\mathrm{m/s}$. Diagonalization of $H$ gives the Dirac energies mentioned above. For two-layer graphene with AB stacking the TB Hamiltonian has the form

$$H = \begin{pmatrix} 0 & v\pi & 0 & 0 \\ v\pi^* & 0 & \gamma_1 & 0 \\ 0 & \gamma_1 & 0 & v\pi^* \\ 0 & 0 & v\pi & 0 \end{pmatrix} \quad (2)$$

where $\gamma_1 = 0.35\,\mathrm{eV}$ is the interlayer hopping energy. Diagonalization of $H$ gives the four energy eigenvalues, $E = \pm[\sqrt{(\gamma_1)^2/4 + (vp)^2} \pm \gamma_1/2]$. Expansion of the square root for the case when $vp \ll \gamma_1$ gives the quadratic dispersion mentioned above with either no gap or a gap of $\pm\gamma_1$ from the symmetry point and with the effective mass in all cases being given by $1/(2m^*) = v^2/\gamma_1$ or $m^* = 0.033 m_0$, where $m_0$ is the free electron mass. The value of $\gamma_1$ is a large energy compared with the thermal energy and the maximum shift of the chemical potential for our experiments.

This TB matrix structure is easily generalized to higher numbers of graphene layers. We assume the usual $ABAB$ stacking structure and are mainly interested in the values of the effective masses and the density of states. The values of the effective masses are listed in column two of Table 1.

| N | $m^*/m_0$ | $g/g_0$ |
|---|---|---|
| 2 | 0.033 | 0.033 |
| 3 | 0.046 | 0.046 |
| 4 | 0.0534 and 0.0204 | 0.074 |
| 5 | 0.0572 and 0.033 | 0.090 |
| 6 | 0.0595, 0.0412 and 0.0147 | 0.115 |

Table 1.   The values of the effective masses $m^*$ relative to the free electron mass $m_0$ and of the two-dimensional densities of states $g$ relative to the value $g_0 = 2m_0/(\pi\hbar^2)$ are shown for different numbers $N$ of graphene layers.   Multiple values of $m^*$ correspond to different parabolic bands.

For three layers the effective mass is enhanced by a factor of $\sqrt{2}$ over the two-layer case.   For four layers the two values of the effective mass are related to the two-layer case by factors of $(\sqrt{5} \pm 1)/2$.   For five layers there are again two effective masses. One is the same as for the two-layer case, and the other one is increased by a factor of $\sqrt{3}$.   Finally, for six layers there are three effective masses.   These values and the resulting two-dimensional density of states $g$ for electrons (or for holes) are collected in Table 1.   The nearest neighbor TB approximation does not give any electron-hole asymmetry or any band overlap.   A further neighbor interaction with strength called $\gamma_4$ does give rise to an electron-hole asymmetry [12,17].

**4 Overlapping-band model**

In order to compare our experiments with the TB theory we first assume a small overlap of the bands, denoted by $E_0$ as shown in the inset of Figure 1. The Dirac bands are not important for the conductance of samples with more than one layer owing to their small density of states. This gives a simple two-band model that is similar to the one proposed by Klein [18] to explain the temperature dependence of the resistance in the parent compound bulk graphite. In the case of graphite this approximation represents an average over the band structure along a corner edge of the three-dimensional Brillouin zone as found in theoretical Slonzcewski–Weiss–McClure [19] model. The main correction that we make to Klein's approximation is to reduce the density of states by a factor of two, because half of the quadratic bands are shifted to higher energies and are not relevant to our low energy measurements.

With band overlap $E_0$ the energy of an electron in the conduction band is $p^2/(2m^*) - E_0/2$ and in the valence band is $E_0/2 - p^2/(2m^*)$. Then using Fermi-Dirac statistics the densities of electrons per area in the conduction band $n_e$ and of holes in the valence band $n_h$ are given by

$$n_e(T) = gk_BT \ln\{1 + \exp[\beta(\frac{E_0}{2} + \mu_F)]\}$$
$$n_h(T) = gk_BT \ln\{1 + \exp[\beta(\frac{E_0}{2} - \mu_F)]\}$$
(3)

where $T$ is the temperature, $k_B$ is the Boltzmann constant, $\beta = 1/(k_BT)$, and $\mu_F$ is the chemical potential. The density of states for electrons and for holes, including a factor of two for spin degeneracy and another factor of two for the K-point degeneracy, is given by

$$g = \frac{2}{\pi\hbar^2}\sum m^*$$
(4)

The resulting values of the density of states are shown in the third column of Table 1.

When a potential difference $V_g$ is applied between the sample and the gate electrode, it induces a charge of $CV_g$ in the sample. An extra electron can either go into the conduction band and increase $n_e$ or go into the valence band and reduce $n_h$. Consequently, with charge equilibrium $n_e = n_h$ for zero gate voltage, the total induced density of charges is

$$\frac{CV_g}{A} = (n_e - n_h)e \qquad (5)$$

Thus equations (3) and (5) give us the following quadratic equation to solve for the fugacity $z = \exp(\beta\mu_F)$ as a function of the gate voltage and temperature.

$$q = \frac{(1+rz)z}{z+r} \qquad (6)$$

where $q = \exp(\frac{\beta CV_g}{geA})$ and $r = \exp(\beta E_0 / 2)$. The solution is

$$z = \left[ q - 1 + \sqrt{(q-1)^2 + 4qr^2} \right] / (2r) \qquad (7)$$

Our conductivity data is then fit to

$$\sigma = e(\mu_e n_e + \mu_h n_h) = egk_B T[\mu_e \ln(1+rz) + \mu_h \ln(1+r/z)] \qquad (8)$$

For numerical accuracy it is convenient to use equation (7) only for positive values of $V_g$ and then to use the fact that $\mu_F(-V_g, T) = -\mu_F(V_g, T)$ for negative values of $V_g$ rather than taking the difference between two large approximately equal numbers. Note that we have ignored the nonuniform charge distribution in the direction perpendicular to the film caused by electrostatic screening [12,20,21].

For data fitting we take $E_0$ and $g$ as free parameters that are independent of temperature. At the charge equilibrium point where $n_e = n_h$ the conductivity is near its minimum, and we have $\mu_F = 0$ at all temperatures. At very low temperatures we have $n_e = n_h = gE_0/2$ and obtain

$$\sigma_{\min}(0) = eg[\mu_e(0) + \mu_h(0)]E_0/2 \qquad (9)$$

while at a finite temperature we have

$$\sigma_{\min}(T) = egk_BT[\mu_e(T) + \mu_h(T)]\ln[1 + \exp(\beta E_0/2)] \qquad (10)$$

When we take the ratio of equation (10) to equation (9) the factor $g$ drops out, thus we can solve for $E_0$. Then equation (9) is used to determine $g$. We initially use our lowest temperature data in equation (9) and our highest temperature data in equation (10) and then make a slight adjustment to obtain the best overall fit to all the data as shown in Figure 2. We find that $E_0 = 16\,\text{meV}$, and the density of states is $g = 0.051g_0$, which is in good agreement with the theoretical value listed in Table 1 for three layers. This leads us to the conclusion that our sample has in fact three layers. Our value of the band overlap is only about half as large as the value 28 meV given in reference [5], which shows that $E_0$ is not an intrinsic property of trilayer graphene. At the maximum gate voltage $V_g = 40\,\text{V}$ the shift of the chemical potential is $\mu_F = 60\,\text{meV}$. Figures 4 and 5 present the results at a low and a high temperature. The minimum conductivity is offset by about 3 volts from zero gate voltage, possibly due to surface contamination of our sample [1,14].

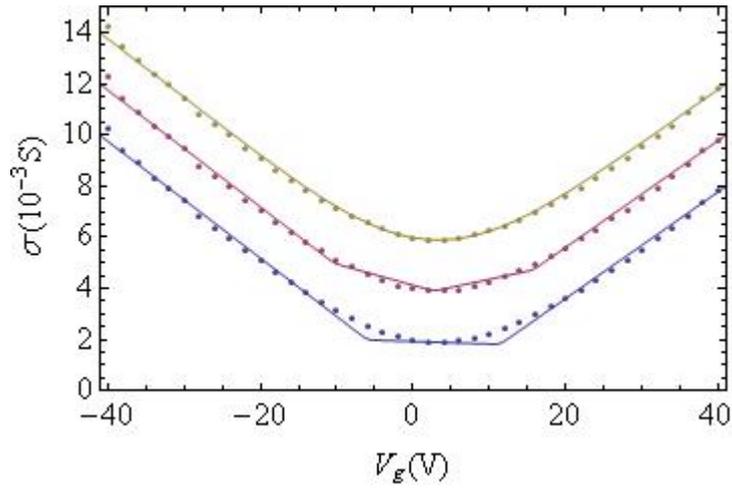

FIG. 4. Conductivity versus gate voltage at $T = 4.5\,\text{K}$. The lowest curve shows the fit to the overlapping-band and two-puddle theories. These two theories give identical results at low temperature. The middle curve shows the fit to the three-puddle theory displaced upward by 2 mS. The top curve shows the fit to the continuous-distribution puddle theory displaced upward by 4 mS. The same data points are shown three times with the corresponding upward displacements.

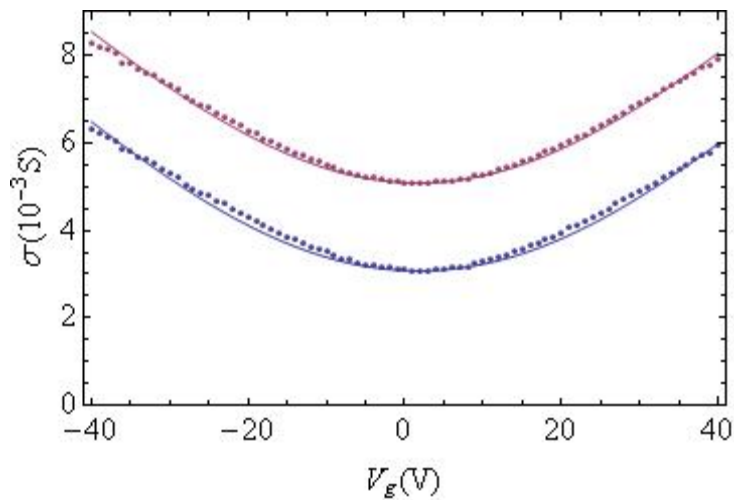

FIG. 5. Conductivity versus gate voltage at $T = 220\,\text{K}$. The bottom curve shows the fit to the overlapping-band theory. The top curve shows the fit to the continuous-distribution puddle theory displaced upward by 2 mS for clarity.

A noticeable discrepancy between the theory and the experimental data is that the theory curve at low temperature in Figure 4 consists of three straight-line segments and is flat near zero gate voltage, unlike the data, which is curved. At zero temperature the theory predicts that the total density of carriers is constant in the band overlap region where $-E_0/2 < \mu_F < E_0/2$. The slight slope shown in Figure 4 is due to the difference between the electron and the hole mobilities. Compared with the work of Morozov *et al.* [22] on bilayer graphene, the shapes of the conductivity versus gate voltage curves are similar.

**5 Puddle models**

The rounding of the experimental data at low temperature near zero gate voltage can originate from spatial inhomogeneity of the charge distribution in the plane of the film. SET and STM work [6-8] has shown that impurities in the substrate cause significantly uneven distributions of charge, called puddles, in the graphene plane. The simplest model for these puddles, which we call the two-puddle model, was introduced by Zhu *et al.* [23]. The energies of the particles are shifted up by a potential $\Delta$ in half of the sample and down by $-\Delta$ in the other half. The band overlap is set equal to zero. Then equation (3) is changed to

$$n_e(T) = \frac{gk_B T}{2}\left(\ln\{1+\exp[\beta(\Delta+\mu_F)]\} + \ln\{1+\exp[\beta(-\Delta+\mu_F)]\}\right)$$
$$n_h(T) = \frac{gk_B T}{2}\left(\ln\{1+\exp[\beta(\Delta-\mu_F)]\} + \ln\{1+\exp[\beta(-\Delta-\mu_F)]\}\right) \quad (11)$$

When we plug equation (11) into equation (5) and solve for $\mu_F$ we obtain a simple linear relationship between the chemical potential and the gate voltage.

$$\mu_F = CV_g / (geA) \quad (12)$$

This result is obtained because the total density of states including both electrons and holes is constant and independent of the band shift $\Delta$ when the band overlap $E_0$ is set equal to zero. This model assumes that the current density is constant in the film and does not take into account the tendency of the current to seek out regions of higher conductivity. We fit the conductivity data to $\sigma = e(\mu_e n_e + \mu_h n_h)$ using equations (11) and (12) and the same values for the mobilities as before. Again we start by taking the ratio of the minimum conductivity at the lowest and highest temperatures to eliminate $g$ and solve for $\Delta$. Then we use the obtained value of $\Delta$ to calculate $g$ from the minimum conductivity at the lowest temperature. As a result we find the values $\Delta = 12$ meV and $g/g_0 = 0.62$. At low temperatures the fit to the data, as shown in the lowest curve in Figure 4, is exactly the same as for the overlapping-band model. It still consists of three straight-line segments, and therefore the two-puddle model also gives an unsatisfactory match to the curvature of the data. At higher temperatures all of the puddle models we consider give the upper curve shown in Figure 5. There is only a small difference between the puddle models and the overlapping-band model here.

All of the sample is not in puddles. Part is not shifted by the impurities in the substrate, so we can consider generalizing the two-puddle model to a three-puddle model where we divide the film into three regions, one shifted by $\Delta$, one shifted by $-\Delta$, and one unshifted. Thus equation (11) is replaced by

$$n_e(T) = \frac{gk_BT}{3} \ln\left(\{1+\exp[\beta(\Delta+\mu_F)]\}\{1+\exp[\beta\mu_F]\}\{1+\exp[\beta(-\Delta+\mu_F)]\}\right)$$
$$n_h(T) = \frac{gk_BT}{3} \ln\left(\{1+\exp[\beta(\Delta-\mu_F)]\}\{1+\exp[-\beta\mu_F]\}\{1+\exp[\beta(-\Delta-\mu_F)]\}\right)$$
(13)

Using equation (12) together with equation (13) and the same fitting method, the new values found for the parameters are $\Delta = 19$ meV and $g/g_0 = 0.60$. At low temperatures the fit to the data is improved as shown by the middle curve in Figure 4. However, the model is still not satisfactory, since it now consists of four straight-line segments and is not curved as in the data.

The best fit that we get to our data is using the continuous-distribution puddle model of Hwang and Das Sarma [10]. A continuous distribution of puddle potentials is taken with a Gaussian distribution.

$$n_e(T) = \frac{gk_BT}{\Delta_0\sqrt{\pi}} \int_{-\infty}^{\infty} d\Delta \exp[-(\Delta/\Delta_0)^2] \ln\{1+\exp[\beta(\Delta+\mu_F)]\}$$
$$n_h(T) = \frac{gk_BT}{\Delta_0\sqrt{\pi}} \int_{-\infty}^{\infty} d\Delta \exp[-(\Delta/\Delta_0)^2] \ln\{1+\exp[\beta(\Delta-\mu_F)]\}$$
(14)

Using equation (12) together with equation (14) we now obtain the parameter values $\Delta_0 = 22$ meV and $g/g_0 = 0.60$. The fit at low temperature shown by the upper curve in Figure 4 is now curved and agrees well with the data. The value of $\Delta_0$ is expected to depend on the particular sample being measured, while the value of $g$ provides a measurement of the average of the electron and hole effective masses.

Our value of $g$ is somewhat higher than the value $g/g_0 = 0.46$ shown in Table 1 for a trilayer film. There are two possible reasons for this discrepancy. Our estimate of the width of our film, which is used to convert resistance to resistance per square $R_\square = Rw/L$, is approximate. On the other hand, the theory we used to construct Table 1 only included nearest-neighbor interactions. The values of the effective masses at the bottom of the band obtained from calculations that include non-nearest-neighbor interactions are somewhat higher. Reference [13] quotes effective masses $m_e^*$ for electrons and $m_h^*$ for holes for bilayer graphene. We have fit the curves shown in Figure 8b of reference [13] to find the effective masses for trilayer graphene. Similarly, we have fit the curves shown in Figure 12 of reference [14] at the bottoms of the bands. The results are listed in Table 2, together with the minimum effective masses from the experimental data of reference [11] for bilayer graphene. Our experimental result lies midway between the two theoretical values shown for trilayer graphene.

| $N$ | $m_e^*/m_0$ | $m_h^*/m_0$ | $g/g_0$ |
|---|---|---|---|
| 2, theory [13] | 0.026 | 0.039 | 0.033 |
| 2, theory [14] | 0.041 | 0.055 | 0.048 |
| 2, experiment [11] | 0.030 | 0.040 | 0.035 |
| 3, theory [13] | 0.044 | 0.066 | 0.055 |
| 3, theory [14] | 0.056 | 0.078 | 0.067 |
| 4, theory [14] | 0.024 and 0.060 | 0.035 and 0.088 | 0.104 |

Table 2. The values of the effective masses $m_e^*$ for electrons and $m_h^*$ for holes relative to the free electron mass are shown for different numbers $N$ of graphene layers. The normalized densities of states $g/g_0$ shown in the last column are the averages of the values shown in columns two and three.

Our result for *g* is obtained only from the temperature dependence of the resistance at the charge-neutral point $\mu_F = 0$. The screening effect does not play a role at this point. When the electron and hole effective masses are not equal the effective value of *g* is an average value of these two masses weighted with the corresponding mobilities.

$$\frac{g}{g_0} = \frac{\mu_e m_e^* + \mu_h m_h^*}{(\mu_e + \mu_h)m_0} \tag{15}$$

Since our electron and hole mobilities are almost equal, it is proper to compare our result for $g/g_0$ with the theoretical results listed in Table 2.

In Figure 2 we demonstrate that the continuous-distribution puddle model agrees well with the data at selected values of $V_g$ over the complete range of temperature. The data fit the model at the highest absolute value of $V_g$ because of the choice of the mobility as a function of temperature shown in Figure 3. The two parameters $\Delta_0$ and *g* are chosen to give the best fit at $V_g = 0$. The most important difference between the continuous-distribution puddle model and the overlapping-band model is shown in the curves at $V_g = -10$ V for low temperatures. The continuous-distribution puddle model matches much better with the data, since this model does not have the unphysical kink that is shown in the lowest curve in Figure 4 for the overlapping-band and two-puddle models in this voltage range.

## 6 Conclusion

In summary, we have performed measurements on trilayer graphene films, illustrating the dependence of the resistance on temperature and gate voltage. A semimetal to metal transition is observed as a function of gate voltage. The data are carefully analyzed according to several theories. We find that the best fit is to the continuous-distribution puddle model. The residual conductivity at low temperature is due to impurities in the substrate, and the impurity potential is distributed according to a Gaussian function. In addition, we determine the average value of the electron and hole effective masses.

[a] rsthom@usc.edu